\documentclass[fleqn]{raa} 
\usepackage{graphicx,times}
\usepackage{natbib}
\usepackage{amssymb,amsmath}
\usepackage{multicol}
\usepackage{float}
\usepackage{makecell}
\usepackage[a4paper]{geometry}

\graphicspath{{./}{figures/}}
\usepackage[a4paper=true,pagebackref=true]{hyperref}

\hypersetup{colorlinks = true, linkcolor = green, anchorcolor = red, citecolor = blue, filecolor = red, pagecolor = red, urlcolor = red}

\begin{document}

   \title{Molecular Clump Extraction Algorithm Based on Local Density Clustering
$^*$
\footnotetext{\small $*$ Supported by the National Natural Science Foundation of China.}
}

 \volnopage{ {\bf 20XX} Vol.\ {\bf X} No. {\bf XX}, 000--000}
   \setcounter{page}{1}

   \author{Xiaoyu Luo\inst{1}, Sheng Zheng\inst{1}, Yao Huang\inst{1}, Shuguang Zeng
      \inst{1}, Xiangyun Zeng\inst{1}, Zhibo Jiang\inst{2}, Zhiwei Chen\inst{2}
   }

   \institute{
   	Center for Astronomy and Space Sciences, China Three Gorges University, Yichang 443000, People's Republic of China {\it xyzeng2018@163.com}\\
      \and
     Purple Mountain Observatory, Chinese Academy of Sciences, Nanjing 210023, People's Republic of China\\
}
\vs \no
   {\small Received 20XX Month Day; accepted 20XX Month Day}

\abstract{
The detection and parametrization of molecular clumps is the first step in studying them. We propose a method based on Local Density Clustering algorithm while physical parameters of those clumps are measured using the Multiple Gaussian Model algorithm. One advantage of applying the Local Density Clustering to the clump detection and segmentation, is the high accuracy under different signal-to-noise levels. The Multiple Gaussian Model is able to deal with overlapping clumps whose parameters can be derived reliably. Using simulation and synthetic data, we have verified that the proposed algorithm could characterize the morphology and flux of molecular clumps accurately. The total flux recovery rate in  $^{13}\rm CO$ (J=1-0) line of M16 is measured as 90.2\%. The detection rate and the completeness limit are 81.7\% and 20 K km s$ ^{-1} $ in $^{13}\rm CO$ (J=1-0) line of M16, respectively. 
\keywords{Molecular Clump, Local Density Clustering, Multiple Gaussian Model}
}

\authorrunning{Xiaoyu Luo. et al. }            %author_head in even pages
\titlerunning{Molecular Clump extraction algorithm based on Local Density Clustering}  % title_head in odd pages

\maketitle

%________________________________________________ sections below
% 
\begin{multicols}{2}
%\twocolumn
\section{Introduction}\label{sec:intro}
The detections of the interstellar molecular hydrogen (H${}_2 $) by \cite{1970ApJ...161L..81C} in the ultraviolet band and carbon (CO) by \cite{1970ApJ...161L..43W} at 2.6 mm, creating an exciting new era in the study of the molecular interstellar medium, while the discovery of organic molecules in the medium led to the birth of molecular astrophysics. As one of the fundamental components of the interstellar medium, molecular clouds consist mainly of molecular gas with a mixture of atoms, ions, dust, and other materials~\citep{2015ARA&A..53..583H,2019RAA....19...17H}. The molecular clouds in the galaxy exhibits the structure over a wide range of scales, from 20 pc or more for giant molecular clouds down to 0.05 pc for dense molecular clumps~\citep{2012The,2013ApJ...779..185K,2020RAA....20..143L}. Modern astronomy proved that the formation of stars is inside the molecular clumps~\citep{2005ApJ...630..250K,2007ARA&A..45..481Z,2009ApJ...699..850K}. Therefore, the molecular clumps are the keys for theoretical models that aim to reproduce the observed characteristics of star formation in the Galaxy~\citep{2017A&A...601A..94R, 2019RAA....19...40T}.

As a consequence, several telescopes (e.g., the FCRAO 14 m, the CfA 1.2 m, the Bell Laboratories 7 m, the PMO 13.7 m telescopes) have devoted to the CO survey projects~\citep{1986ApJS...60....1S,2001ApJ...547..792D,2001ApJS..136..137L,2011ChAA..35..439Z}. These CO surveys will lead to a better understanding of the evolution of molecular clump, the initial mass function of stars, as well as the structure and dynamic evolution of the Milky Way~\citep{2015ARA&A..53..583H}. With the progresses of the CO survey, it is impractical to process massive amounts of data manually. Therefore, a stable and reliable algorithm for automatically detecting the molecular clumps has become the focus. Several algorithms have been used to detect molecular clumps, such as GaussClumps, FellWalker, ClumpFind and Reinhold~\citep{1990ApJ...356..513S,2015A&C....10...22B,1994ApJ...428..693W,2007ASPC..376..425B}. GaussClumps was first applied to the M17 molecular cloud to detect molecular clumps, and then frequently applied to the detection of clumps in other molecular clouds~\citep{1998A&A...338..262S,2009MNRAS.395.1805D,2009MNRAS.395.1021L}. The ClumpFind algorithm was applied to the detection of compact structures in the Rosette molecular clouds. A new giant filament was found by \cite{2016RAA....16...56Z} with a statistical study on the giant molecular cloud M16~\citep{2002ApJ...565L..25S}.

Studies shows that the ClumpFind is very sensitive to the initial parameters, and the GaussClumps can only fit a strict elliptic shape. FellWalker exhibit the best performance in detection completeness and parametrization~\citep{2020RAA....20...31L}. However, it should be noted that the GaussClumps and the ClumpFind algorithms are affected by the initial parameters, and the algorithms themselves are designed to simulate the ``human eye" for molecular clump recognition, which have certain limitations~\citep{2008ApJ...679.1338R}. Moreover, for large amounts of molecular cloud data, it is clearly not feasible to rely on algorithms with repeatedly setting parameters by users, although it is possible to achieve satisfactory detection results in certain cases. Therefore, we need to design an algorithm which has fewer parameters or can be adjusted more easily based on the physical properties.

One of the domainant features of molecular clumps with increased local intensity and different shapes is that they are embedded in molecular gas of lower average bulk density~\citep{1986ApJ...300L..89B,1992ApJ...393L..25L}. The Local Density Clustering (LDC) algorithm~\citep{2014Science...344..6191} has its basis on assumptions that the cluster centers are surrounded by neighbors with lower local density and they are at a relatively large distance from the points with a higher local density, which is similar to the characteristics of melecular clumps. Therefore, we attempt to adopt the LDC in the detection of molecular clumps. In Section \ref{sec:algorithm introduction}, the molecular clump detection algorithm based on LDC and parametrization based on Multiple Gaussian Model (MGM) are introduced. In Section~$ \ref{sect:compare} $, the 3D simulated datasets with different number density are described. The performance of the LDC \& MGM is compared with traditional algorithms on the datasets. The investigation of the completeness and parametrization of the algorithm in real melocular clump data are in Section~$ \ref{sect:Experiment in real data}$, while the summary is in Section~$ \ref{conclusion}$.
%which is consist with the structure of molecular clumps.

\section{Algorithms}\label{sec:algorithm introduction}
\subsection{The LDC Algorithm} \label{subsec:The local density cluster}
\subsubsection{Features Extraction} \label{subsec:extract features}
The algorithm firstly compute three parameters of a point: the local density, the distance, and the gradient. The local density $\rho_i$ of a point $p_i$ is defined as:
\begin{equation}
\label{Formula 1}
\rho_{i}=\sum_{j} (e^{-\left(d_{i j} / d_{c}\right)^{2}} \cdot I_{j})
\end{equation}
where the $d_{c}$ represents the cut-off distance, $d_{i j}$ represents the distance between $p_{i}$ and $p_{j}$, $I_{j}$ represents the intensity at $p_{j}$. 

The distance $\delta_{i}$ of a $p_i$ is defined as:
\begin{equation}
\label{Formula 2}
\begin{array}{l}
\delta_{i}=\min \limits_{j: \rho_{j}>\rho_{i}} {d}_{i j}  

\end{array}
\end{equation}
$ \delta_{i} $ is measured by computing the minimum distance between $p_{i}$ and any other point with higher density. Specially, $\delta_{i}$ is set to be the maximum $ \delta $ if $p_i$ with highest density. The distances $\delta_{i}$ are normalized.

The nearest route could be obtained while calculating the distances. The index of the point with the smallest distance among all the data points whose density is greater than the current $p_i$ is recorded in the vector $ N^{(p)} = \{n_{1}^{(p)},n_{2}^{(p)},\cdots,n_{i}^{(p)},\cdots,n_{n}^{(p)}\} $:
\begin{equation}
\label{Formula 3}
n_{i}^{(p)}=\left\{\begin{array}{l}
0, \text { if } \delta_{i} = \text{max}(\delta_{i}) \\
j, \text { if } \delta_{i}=d_{i j} 
\end{array}\right.
\end{equation}
among them, $ n $ represents the total number of data points, the point with longest distance is set as 0 in the vector $ N^{(p)} $.

The gradient $\nabla_{i}$ is defined as:
\begin{equation}
\label{Formula 4}
\nabla_{i}= \frac{\rho_{j}-\rho_{i}}{\delta_{i}}
\end{equation}
where $ \rho_{j} $ and $ \delta_{i} $ are defined in Formula~$ (\ref{Formula 2}) $.
\subsubsection{The Clump Center Determination} \label{subsec:Decision Graph}
\begin{figure}[H]
	\centering
	\includegraphics[angle=0,width=86mm]{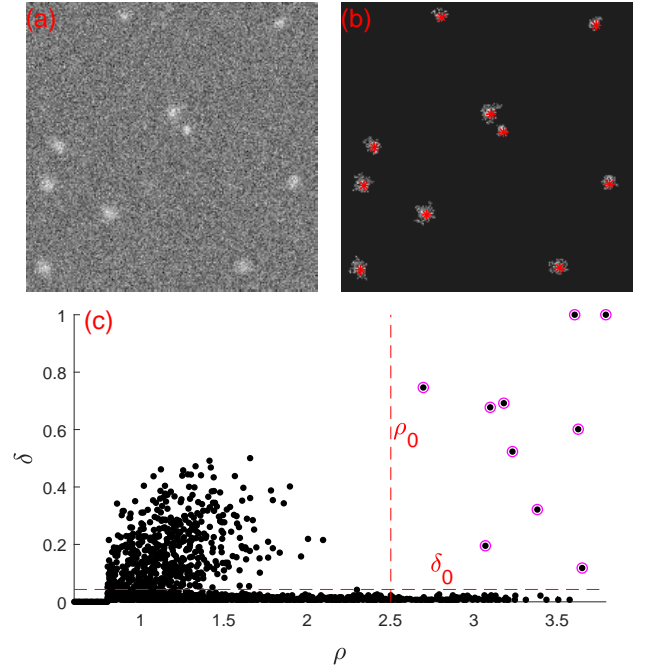}
	\caption{The example of algorithm on 2D simulated data. (a) The 2D data contains 10 simulated clumps. (b) Clustering result. (c) Decision Graph of the data in (a).} 
	\label{Decision}
\end{figure}

After calculating three parameters, as shown in Figure~\ref{Decision}, the distance $ \delta $ is plotted against the density $ \rho $, which is referred to the Decision Graph. The simulated data with 10 clumps is shown in Figure~\ref{Decision} (a), while the detected clumps are shown in Figure~\ref{Decision} (b) with centers marked by red stars. Figure~\ref{Decision} (c) shows the Decision Graph, where the centers of the detected clumps are marked with circles.  Whether $ p_i $ is the center point of a clump or not is judged by:
\begin{equation}
\label{Formula 5}
p_{i}=\left\{\begin{array}{l}
p_{k}^{(C)}, \text{ if } \delta_{i} \geq \delta_{0}, \rho_{i} \geq \rho_{0} \\
p^{(n o n-C)}, \text{ else }
\end{array} \quad i=1,2, \cdots, n\right.
\end{equation}
where $ p_{k}^{(C)} $ represents the  center point of the $ k_{th} $ clump, $p^{(non-C)}$ represents the  point of non-clump. $\delta_{0}$ and $ \rho_{0} $ are hyper-parameters of our algorithm, where $\delta_{0}$ represents the minimum distance between the centers of the two clumps, and $ \rho_{0} $ represents the minimum peak intensity value of a candidate clump.
\subsubsection{Route Clustering}
\label{clustering}
According to the information recorded in the route vector $ N^{(p)} $, the route $ P_k $ end with the $ k_{th} $ center point $ p_{n_k}^{(C)} $ of clump can be obtained:
\begin{equation}
\label{Formula 6}
P_{k}=\left\{p_{1}^{(k)}, p_{2}^{(k)}, \cdots, p_{j}^{(k)},\cdots, p_{n_k}^{(C)}\right\} \text { }k=1,2, \cdots, N
\end{equation}
other points of non-clump are divided into the $ k_{th} $ clump $ C_{k} $ according to route $ P_k $:
\begin{equation}
\label{Formula 7}
p_{j} \in C_{k}, \text { if }  p_{j} \in P_{k} \text { }j=1,2, \cdots, n_k
\end{equation}
where $ N $ and $n_{k}$ represent the number of clumps and the number of data points in the clump $C_{k}$, respectively.
\subsubsection{Clump Region Determination}
\label{gradient}
The region of the individual clump $ C_{k} $ can be determined according to $ \rho $ and $ \nabla $:
\begin{equation}
\label{Formula 8}
p_{j}^{(k)} \in C_{k}, \text { if } \rho_{j} \geq  \bar{\rho}  \text { or } \nabla_{j} \geq \nabla_{0}  \text{ }j=1,2, \cdots, n_{k}
\end{equation}
where $ p_{j}^{(k)} $ represents the $ j_{th} $ point in clump $ C_{k} $, $ \bar{\rho} $ is the average density of the clump $ C_{k} $, $ \nabla_{0} $ is the hyper-parameter. The individual clump could be segregated as $ \rho_j $ greater than $ \bar{\rho} $ or $ \nabla_j $ greater than $ \nabla_{0} $. The morphological image processing is employed to fill in holes among detected clumps and to smooth its boundary.

\subsubsection{False Clumps Exclusion}
\label{determine number of cluster}
The isolated noise points with high peak intensity value could be recognized as false clumps. The smallest clump should have enough data points to form it. Therefore, the false detected clumps could be eliminated by the following criteria:
\begin{equation}
\label{Formula 9}
\text {C}_{k}=\left\{\begin{array}{l}
\text {True, if } n_{k} \geq n_{0} \\
\text {False, else} 
\end{array}\right.
\end{equation}
where $n_{0}$ is the minimum data point number of a clump.
\subsubsection{Clump Characterization}
\label{outcat}
The algorithm will provide a pixel mask which is the same shape as the supplied data array. In the mask, the pixel points belonging to the same clumps are marked with an integer, while points that are not assigned are marked with -1. Finally, a table in which each row describes an individual clump is obtained. In each column of the table, $Peak_i$ and $Cen_i$ represent the position of the clump peak intensity value and centroid on axis $i$ ($ i =1, 2, 3 $), respectively. $Size_i$ represents the size of the clump on axis $i$. $Sum$ and $Peak$ represent the total flux and peak intensity value of the clump, respectively. The definition of the centroid is as follows:
\begin{equation}
\label{Formula 10}
Cen_{i}=\frac{\sum_{j=1}^{n_k} (I_{j} \cdot x_{j})}{\sum_{j=1}^{n_k} I_{j}}
\end{equation}
the $ Size_i $ of the clump $ C_k $ on axis $ i $ is defined as:
\begin{equation}
\label{Formula 11}
Size_i =\sqrt{\frac{\sum_{j=1}^{n_k} (I_{j} \cdot x_{j}^{2})}{\sum_{j=1}^{n_k} I_{j}}-\left(\frac{\sum_{j=1}^{n_k} (I_{j} \cdot x_{j})}{\sum_{j=1}^{n_k} I_{j}}\right)^{2}}
\end{equation}
where $I_{j}$ and $x_j$ represent the intensity and position of $p_j$, respectively. For the clump with a Gaussian profile, the size is equal to the standard deviation of the Gaussian.
\subsubsection{\textbf{Algorithm Summarising}}
\label{summarising}
The input of algorithm is a 3D (or 2D) data array. $\delta_{0}$ and $ \rho_{0} $ are key hyper-parameters of the algorithm, where $\delta_{0}$ represents the minimum distance between the centers of the two clumps, and $ \rho_{0} $ represents the minimum peak intensity value of a candidate clump. $n_{0}$ is the minimum data point number of a clump and $ \nabla_{0} $ is used to determine the region of a clump. The local density of a point is calculated with the cut-off distance ($d_{c}$). The input and parameters of the LDC algorithm are listed in Table~\ref{input_parameters}. The outputs include masks indicating the pixels that contribute to each clump, and catalogs holding clump positions, sizes, peak values and total fluxes. The output and parameters of the LDC algorithm are listed in Table~\ref{output_parameters}.
\begin{table}[H]	
	\centering 	
	\caption[]{The input and parameters of LDC algorithm}
	\label{input_parameters}

	\begin{tabular}{clcl}
		\hline\noalign{\smallskip}
		\makecell[c]{Input}& \makecell[c]{Description}&
		\makecell[c]{Default value} \\
		\hline\noalign{\smallskip}
		
			\makecell[c]{Data array} &\makecell[c]{3D or 2D data array}  & \makecell[c]{}\\

		\hline\noalign{\smallskip}
		
		\makecell[l]{Parameters}
		&\makecell[c]{$\rho_{0}$\\$\delta_{0}$\\$ \nabla_{0}$ \\$n_{0}$ \\ $ d_c$ }
		&\makecell[c]{  $3 \sigma $\\4\\  0.01 \\  27 \\ 0.8 } \\
		
		\noalign{\smallskip}\hline
		
	\end{tabular}
	
\end{table}

\begin{table}[H]	
	\centering 	
	\caption[]{The output of LDC algorithm}\label{output_parameters}
	
	\begin{tabular}{clcl}
		\hline\noalign{\smallskip}
		\makecell[c]{Output}& \makecell[c]{Description}& \makecell[c]{Explanation}       \\
		\hline\noalign{\smallskip}
		
		\makecell[c]{Mask}&\makecell[c]{Data array}  & \makecell[c]{The same shape as the input data}\\

		\hline\noalign{\smallskip}
		
		\makecell[l]{Parameters}&\makecell[c]{$Peak_i$  \\($ i=1,2,3 $)\\$Cen_i$  \\($ i=1,2,3 $)\\$Size_i$ \\ ($ i=1,2,3 $) \\$Peak$ \\ $ Sum$  } & \makecell[c]{The position of the clump\\ peak value on axis $ i $\\The position of the clump\\ centroid on axis $ i $\\The size of the clump\\ on axis $ i $ \\The peak value of the clump \\ The total flux of the clump }\\
		
		\noalign{\smallskip}\hline
		
	\end{tabular}
	
\end{table}
The detection of the algorithm is not affected by neither the shape of the clumps nor the dimensionality of the space they embedded in. The detection results of the LDC in defferent number density and different PSNR are shown in the Table~\ref{tab1}. The size of a simulated data is $ 100\times100\times100 $.

\begin{table}[H]
	
	\centering 

		\caption[]{The performance of the algorithm}\label{tab1}
		
		%%Please Capitalize the First Letter of Each Notional Word in table's caption
		
		\begin{tabular}{clcl}
			\hline\noalign{\smallskip}
			Number density levels &  \makecell[c]{High}      & \makecell[c]{Medium} & \makecell[c]{Low}                   \\
			\hline\noalign{\smallskip}

			\makecell[c]{Number of clumps in \\$ 100\times100\times100 $ data array} & \makecell[c]{100} & \makecell[c]{25}     & \makecell[c]{10}  \\ % new variable
			Recall rate (PSNR $ \geq $ 6)  & $ >80\%  $    &   $ >91\% $     & $ >97\% $                  \\
			Precision rate (PSNR $ \geq $ 6)  & $ >90\% $     &   $ >96\% $     & $ >98\% $                  \\
			$ F_1 $ (PSNR $ \geq $ 6)  & $ >86\% $     &   $ >94\% $     & $ >98\% $ \\
			\noalign{\smallskip}\hline
		\end{tabular}

\end{table}

\subsection{Parametrization Based on MGM}\label{sect:parameter restoration}
Traditional algorithms are used to segregate overlapping molecular clumps, and there could be large deviation in the parameter estimation of overlapping molecular clumps. Therefore, we adopt MGM to realize the parametrization in this paper.
%Inspired from the problem can be effectively solved by the Gaussian Mixture Model~\cite{2008Central}, we adopt MGM as the basis of parametrization in our proposed algorithm.
%we adopt MGM to realize the parametrization in our proposed algorithm.
\subsubsection{The 3D Gaussian Model}
\label{3D Gaussian model}
The observation data of molecular clump is a 3D data array. The first and second dimension of the 3D data array stand for the galactic longtitude and latitude, respectively. The third dimension of the 3D data array stands for
the velocity. Due to the spatial and velocity are not related, the tilt angles of simulated clumps only appear on the galactic longtitude - latitude plane. 
Therefore, the 3D Gaussian Distribution is described as:
\begin{equation}\label{3d rotate gaussian model}
\begin{gathered}
f(x, y, v)=A \exp \left\{-\left[\left(\frac{\cos ^{2} \theta}{2 \sigma_{x}^{2}}+\frac{\sin ^{2} \theta}{2 \sigma_{y}^{2}}\right)\left(x-x_{0}\right)^{2}\right.\right. \\
+\left(\frac{\sin 2 \theta}{2 \sigma_{y}^{2}}-\frac{\sin 2 \theta}{2 \sigma_{x}^{2}}\right)\left(x-x_{0}\right)\left(y-y_{0}\right) \\
\left.\left.+\left(\frac{\sin ^{2} \theta}{2 \sigma_{x}^{2}}+\frac{\cos ^{2} \theta}{2 \sigma_{y}^{2}}\right)\left(y-y_{0}\right)^{2}+\frac{\left(v-v_{0}\right)^{2}}{2 \sigma_{v}^{2}}\right]\right\}
\end{gathered}
\end{equation}
where $(x_{0}, y_{0}, v_0)$ represents center point of the distribution, $\sigma_{x}, \sigma_{y}, \sigma_{v}$ represent the standard deviations in the three axes, respectively. The variable $ A $ represents amplitude of the distribution, and $\theta$ represents the tilt angle on the $x-y$ plane.
\subsubsection{The 3D MGM}
\label{3D Gaussian Mixture Model}
\begin{figure}[H]
	\centering
	\includegraphics[angle=0,width=86mm]{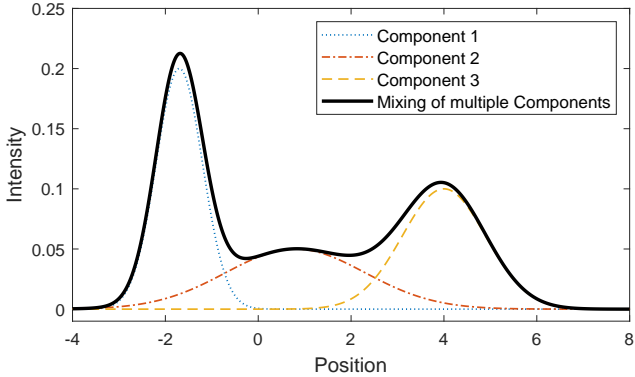}
	\caption{Overlapping of three Gaussian components. The black line is composed of a combination of three gaussian distributions. Dashed lines represent each gaussian components.} 
	\label{three component gaussain}
\end{figure}

For the scenario where multiple gaussian components overlap, adopting a single gaussian distribution to explain will lead to serious deviation. Taking Figure~$ \ref{three component gaussain} $ as an example, the black solid line represents the actual data obtained, and the three dashed lines represent the actual components. The MGM can effectively solve this problem.
The MGM is defined as follows:
\begin{equation}
f_{mix}(x, y, v ; \psi)=\sum_{k=1}^{K} f_{k}\left(x, y, v ; \psi_{k}\right)
\end{equation}
where $ f_{k}(x, y, v ; \psi_{k}) $ represents $ k_{th} $ 3D Gaussian Distribution described in Formula~($ \ref{3d rotate gaussian model} $), and $ K $ is the number of Gaussian components. $\psi=\left\{\psi_{1}, \psi_{2},  \cdots, \psi_{k}, \cdots \psi_{K} \right\}$ defines the parameters of the model. $\psi_{k}$  represents the prarmeters of the $ k_{th} $ Gaussian Distribution. $\psi_{k}$ is specified as :
\begin{equation}
\psi_{k}=\left\{A^{(k)}, x_{0}^{(k)}, \sigma_{x}^{(k)}, y_{0}^{(k)}, \sigma_{y}^{(k)}, \theta^{(k)}, v_{0}^{(k)}, \sigma_{v}^{(k)}\right\}^{T}
\end{equation}
Using the LDC algorithm described in Section~\ref{subsec:The local density cluster}, a series of clumps $ (C_1, C_2, \cdots, C_k, \cdots, C_N) $ can be obtained, and the parameters of those clumps calculated in Section~$ \ref{outcat} $ could serve as the initial value $(\psi_{1}^{(0)}, \psi_{2}^{(0)}, \cdots, \psi_{k}^{(0)}, \cdots, \psi_{N}^{(0)}) $ of the model. The clump $ C_i $ and $ C_j $ are considered to overlap each other when $ \Vert (x_{i0}-x_{j0}, y_{i0}-y_{j0}, v_{i0}-v_{j0}) \Vert_2 \leq \Vert (\sigma_{ix}+\sigma_{jx}, \sigma_{iy}+\sigma_{jy}, \sigma_{iv}+\sigma_{jv}) \Vert_2$ ($ \Vert \cdot \Vert_2$ represents two-norm). Then the parameters $(\psi_{1}^{(0)}, \psi_{2}^{(0)}, \cdots, \psi_{k}^{(0)}, \cdots, \psi_{m}^{(0)}) $  of the overlapping clumps (suppose the number of overlapping clumps is $ m $) could serve as the initial parameters while using the MGM fit those overlapping clumps. Finally, the catalogue with various clump parameters will be obtained via MGM fitting method.

\section{Comparison with other Algorithms}\label{sect:compare}
\subsection{Detection Accuracy}
\label{compare detect acc}
\subsubsection{3D Simulated Data}
\label{3d simulation data}
\begin{figure}[H]
	\centering
	\includegraphics[angle=0,width=86mm]{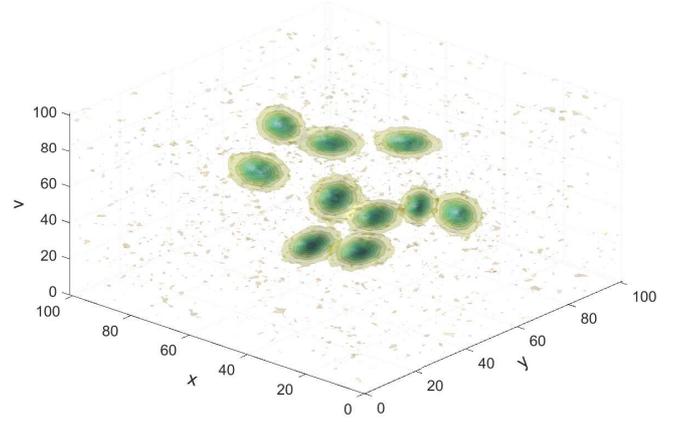}
	\caption{3D display of simulated clumps.} 
	\label{3D display of one simulated data array}
\end{figure}

The simulated datasets are composed of different number density data with the size of $ 100\times100\times100 $, and data at low, medium and high density contain 10, 25 and 100 simulated clumps, respectively. The peak intensity value of the clump take values from 2 to 10, while the size of the clump in velocity axis take values from 3 to 5 and the spital size in the $ x $ and $ y $ axes take values from 2 to 4 ($ \rm{FWHM} = 2.35\times size $). The tilt angles of the simulated clumps on the $ x-y $ plane vary from $0 ^{\circ} $ to $180 ^{\circ} $. Gaussian noise is added to the simulated clumps with a root-mean-square $ (rms) $ of 1. For each number density, we generate a total of 10000 simulated clumps. Figure~$ \ref{3D display of one simulated data array} $ shows the 3D display of one simulated data array.
\subsubsection{Detection based on LDC Algorithm}
\label{detect data}
\begin{figure}[H]
	\centering
	\includegraphics[angle=0,width=86mm]{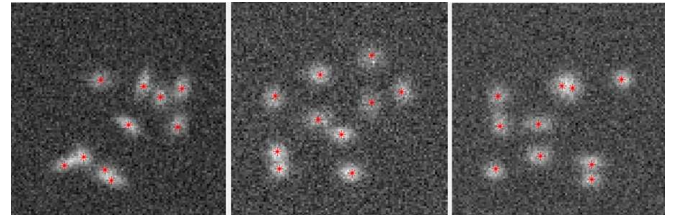}
	\caption{The centers of detected clumps are marked on the intergrated intensity maps with red asterisks. From left to right are integral maps of $ x-y $, $ x-v $ and $ y-v $ planes, respectively.
	} 
	\label{fig2-3}
\end{figure}

As shown in Figure~$ \ref{fig2-3} $, from left to right are the integral maps on the three planes of $ x-y $, $ x-v $ and $ y-v $, respectively. The center points of clumps are marked with red asterisks on the integral graphas.

\begin{figure}[H]
	\centering
	\includegraphics[angle=0,width=86mm]{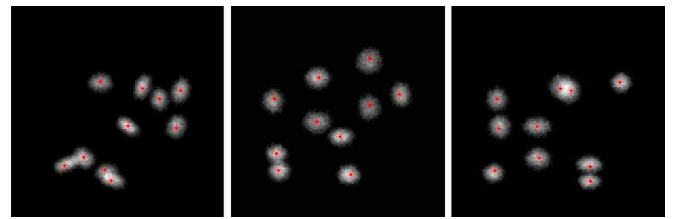}
	\caption{The intergrated intensity maps of detected  clumps are marked with red asterisks. From left to right are integral maps of the $ x-y $, $ x-v $ and $ y-v $ planes, respectively.
	}
	\label{fig2-4}
\end{figure}

Combined with the $ \nabla $ and $ \rho $ of each data point, the members and region of the clump $ C_k $ could be determined by Formula~($ \ref{Formula 8}$). (1) Using $ \nabla_{0} $ as the threshold, the point set $ A_{1} $ with $ \nabla $ greater than $ \nabla_{0} $ is the main part of the clump $ C_k $. (2) The average density $ \bar{\rho} $ is calculated based on the point set $ A_{1} $, then the point set $ A_{2} $ represents $ \rho $ greater than $ \bar{\rho} $ which is also the main part of the clump. (3) The union of $ A_{1} $ and $ A_{2} $ could form the region of an individual clump $ C_k $. The detection results are shown in Figure~$ \ref{fig2-4} $. The region of each clump will determined while the false clump will eliminated by Formula~$ (\ref{Formula 9}) $. Finally, the parameter estimation in Section~\ref{outcat} will be performed.
\subsubsection{Evaluation Indicators}
\label{detect evaluation}
\begin{figure*}[ht]
	\centering
	\includegraphics[angle=0,width=172mm]{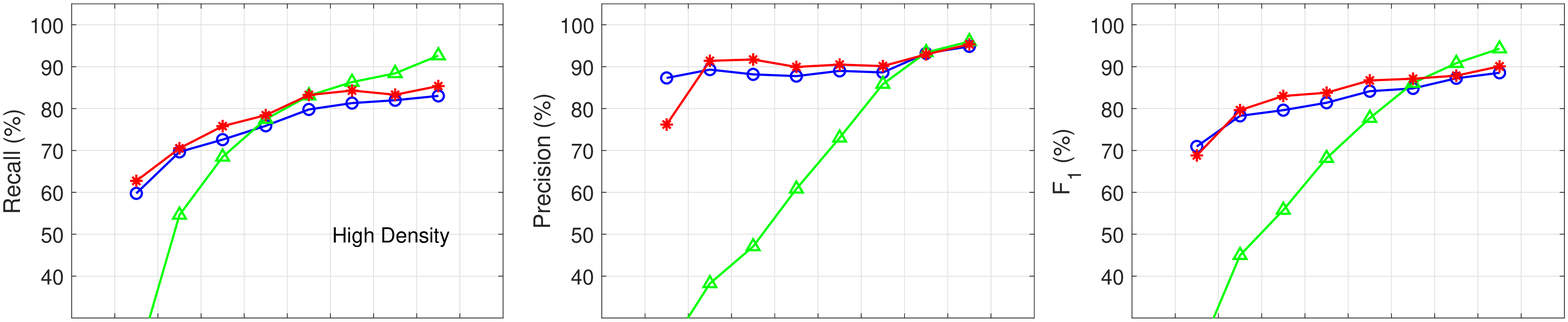}
	\includegraphics[angle=0,width=172mm]{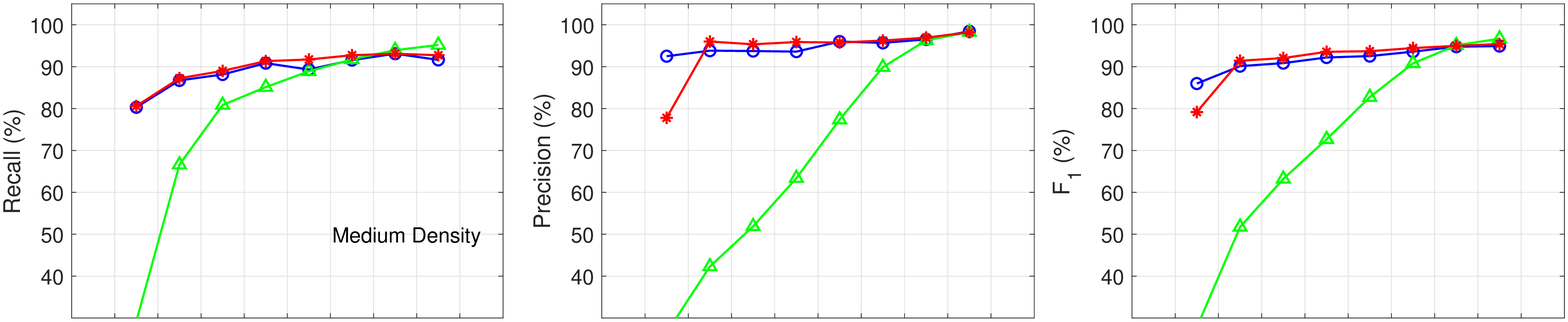}
	\includegraphics[angle=0,width=172mm]{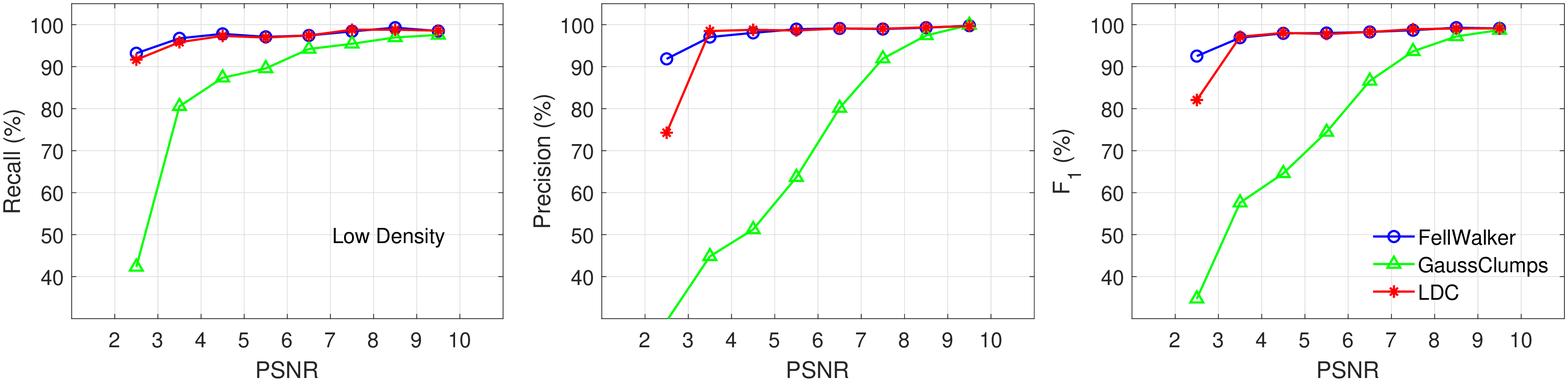}
	\caption{
		The evaluation indicators $R$, $P$ and $F_1$ of the three algorithms are plotted against the PSNR at different number density. \textit{Top panel}: the detection statistics of the three algorithms in high density, from left to right are $R$, $P$ and $F_1$, respectively. \textit{Middle panel}: same as above but for medium density. \textit{Bottom panel}: same as above but for low density.} 	
	\label{fig_recall_precision_F1}
\end{figure*}
The detection of molecular clumps is considered to be correct if the euclidean distance between the center of the detected clump and the center of the simulated is less than 2 pixels in the three axes.

Four statistics are obtained by the detection results as follows: True-Positive ($ TP $), True-Negative ($ TN $), False-Positive ($ FP $), and False-Negative ($ FN $)~\citep{3DCNN}. The evaluation indicators for the algorithm include: $ recall $   $rate$ ($ R $), $ precision$   $rate$ ($ P $) and comprehensive score ($F_{1} $). The $ R $, $ P $ and $ F_{1} $ are defined as: 
\begin{equation}
\label{recall,precision,f1}
R=\frac{T P}{T P+F N}, P=\frac{T P}{T P+F P}, F_{1}=\frac{2 \cdot P \cdot R}{P+R}
\end{equation} The accuracy and completeness of detection are reflected in $ P $ and $ R $, respectively. A good detection algorithm should have higher $ P $ and $ R $. Usually the two indicators will show the opposite trend. The comprehensive performance ability of the algorithm is mainly reflected in $ F_{1} $. 
\subsubsection{Detection Comparison}
\label{Testing performance evaluation}
The GaussClumps, FellWalker and LDC are employed in the detection of simulated clumps. Figure~$ \ref{fig_recall_precision_F1} $ shows the evaluation of indicators $ R $, $ P $ and $F_{1} $ of the GaussClumps, FellWalker, and LDC algorithms in different peak signal-to-noise ratio (PSNR) levels and different density. The PSNR is defined as the ratio of the peak intensity value of the simulated clump to the $ rms $ of noise. As the PSNR decreases, $ R $ of these algorithms at different density levels starts to decrease, especially when the PSNR is less than 4. The FellWalker and LDC algorithms generally have high $ P $, while the same indicator of GaussClumps performed worse with the PSNR less than 6. It is obvious that $ R $ of those algorithms at different density hold high level with the PSNR above 6, and $ P $ of those algolrithms hold high level with the PSNR greater than 7, while $ P $ of GaussClumps gradually descend with the decrease of the PSNR.

The top panel in Figure~$ \ref{fig_recall_precision_F1} $ shows the $ R $, $ P $ and $F_{1} $ of GaussClumps, FellWalker, and LDC algorithms at high density from left to right, respectively. The $ R $ and $ P $ are above 80\% for those algorithms when the PSNR is greater than 7. While $ P $ of GaussClumps is greater than FellWalker and LDC in the case of the high PSNR, and $ R $ of GaussClumps is lower than the two algorithms in low PSNR. For those clumps in the simulation that overlap heavily or even merge into new clumps, FellWalker and LDC are unable to distinguish these clumps, leading to a decrease in $ R $. Since Gaussclumps detects the clumps by fitting, it can separate the overlapping clumps from each other, thus improving $ R $. The middle panel shows the same as above but for medium number density. $ P $ of the three algorithms are essentially the same as in the case of high density, but $ R $ of those algorithms have increased and the gap between GaussClumps and the other two algorithms is further reduced with the PSNR above 6. The bottom panel shows the same as above but for low number density. $ P $ of the three algorithms are basically the same as in the case of high density, but $ R $ are above 90\% for the three algorithms when the PSNR is greater than 5, and $ R $ of GaussClumps is lower than the other two algorithms. \par
The experimental results show that $ P $ and $ R $ of FellWalker and LDC algorithms can be maintained at high level, but $ R $ decreases in the case of high density. GaussClumps algorithm has high $ R $ and $ P $ at the certain PSNR indicating that it is susceptible to noise. In terms of the comprehensive performance indicator $F_1$, the FellWalker and the LDC algorithm are essentially the same, both outperforming the GaussClumps algorithm in low PSNR.

\subsection{Parametrization}
\label{extracting parameters}
\subsubsection{Evaluation Indicators}
\label{evaluation index para}
To investigate the performance of the algorithm in terms of parametrization accuracy, various measured pararmeters are compared with their input values, peak intensity, total flux, tilt angle, size, and position of the clump. For each parameter, the absolute deviation of the position $ E(\Delta X) $, angle $ E(\Delta \theta) $, and the relative deviation of size $ E(\Delta S) $, peak intensity $ E(\Delta I) $ and total flux $ E(\Delta F) $ are calculated. Those evaluation indicators are defined as:
\begin{equation}
E(\Delta X)=\frac{1}{N} \sum_{i=1}^{N} (X_{i}^{(s)}-X_{i}^{(m)})
\end{equation}
\begin{equation}
E(\Delta \theta)=\frac{1}{N} \sum_{i=1}^{N}  ({\theta }_{i}^{(s)}- {\theta }_{i}^{(m)})
\end{equation}
\begin{equation}
E(\Delta S)=\frac{1}{N} \sum_{i=1}^{N} \frac{ S_{i}^{(s)}-S_{i}^{(m)}}{S_{i}^{(s)}}
\end{equation}
\begin{equation}
\label{d_peak}
E(\Delta I)=\frac{1}{N} \sum_{i=1}^{N} \frac{ I_{i}^{(s)}- I_{i}^{(m)}}{I_{i}^{(s)}}
\end{equation}
\begin{equation}
\label{d_flux}
E(\Delta F)=\frac{1}{N} \sum_{i=1}^{N} \frac{ F_{i}^{(s)}- F_{i}^{(m)}}{F_{i}^{(s)}}
\end{equation}
where $ N $ represents the number of simulated molecular clumps which are detected correctly by the algorithm. The superscript $ s $ and $ m $ represent the parameters of the simulated molecular clumps and measured by the algorithm, respectively. $ X $, $ S $, $ I $ and $ F $ represent the position, size, peak and total flux of the clump, $ \theta $ represents the tilt anlge on the $ x-y $ plane of the clump.
\subsubsection{Performance}
\label{Testing performance para}
We launched statistical experiments to compare the parametrization performance of FellWalker, GaussClumps and LDC \& MGM algorithms. The high density simulated data described in Section~\ref{3d simulation data} are used in the statistical experiments.

\begin{figure}[H]
	\centering
	\includegraphics[angle=0,width=86mm]{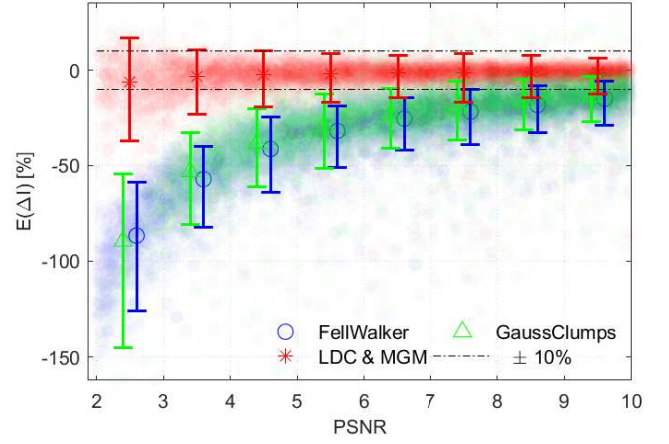}
	\caption{The statistics of relative deviation in peak intensity by the three algorithms as a function of the PSNR. The blue, green and red dots show the distribution of the individual measurements. The Special symbols and error bars represent the median and standard deviation of accuracy, respectively. Two dashed horizontal lines represent the relative deviation of $ \pm10\% $.} 	
	\label{peak}
\end{figure}

Figure~$ \ref{peak} $ shows the relative deviation of peak intensity value as a function of the PSNR. The vertical axis represents the relative deviation of peak intensity between the simulated clump and measured clump, and the horizontal axis represents the PSNR of the clump. The blue, green and red dots represent the relative deviation of clumps detected by FellWalker, GaussClumps and LDC \& MGM, respectively. When the dot is above 0, it means that the value of measured by algorithm is less than the simulated, otherwise, the value of measured is greater than the simulated. Error bars represent standard deviation of accuracy. The blue circle, green triangle and red asterisk represent the median of relative deviation measured by the FellWalker, GaussClumps and LDC \& MGM algorithm, respectively.
From Figure~$ \ref{peak} $ we can see that as the PSNR of the simulated clump increase, the deviation of the GaussClumps and FellWalker algorithms gradually decrease, while the peak intensity values measured by both algorithms are greater than the simualted. The deviations obtained by the LDC \& MGM algorithm are close to 0 with the dispersion decreased gradually, indicating that the peak intensities estimated from the LDC \& MGM algorithm is more reliable.
\begin{figure}[H]
	\centering
	\includegraphics[angle=0,width=86mm]{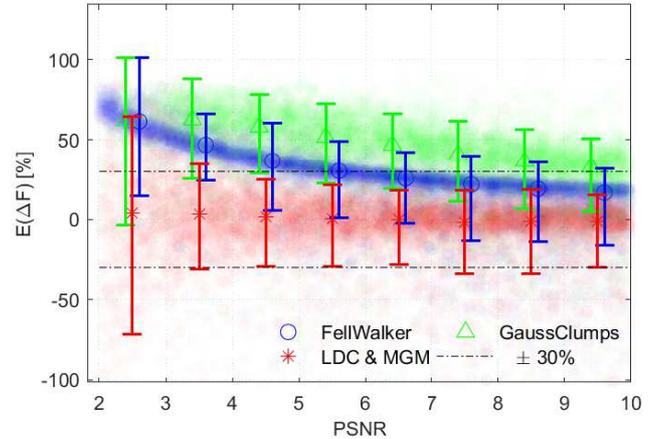}
	\caption{The statistics of the relative deviation in the total flux as a function of the PSNR. Two dashed horizontal lines represent the relative deviation of $ \pm30\% $. The blue, green and red dots, special symbols and error bars have the same meaning as Figure~\ref{peak}.} 
	\label{flux}
\end{figure}
The total flux is an important parameter, which is directly related to the column density and mass of a molecular clump. As can be seen in Figure~$ \ref{flux} $, the total fluxes of GaussClumps and FellWalker algorithms are smaller than the simulated values. The reason is that both algorithms have a cutoff threshold for background noise in detecting molecular clumps and can only detect part of clumps. The most deviation of LDC \& MGM does not exceed $ \pm $30\% with the PSNR greater than 4, indicating that the LDC \& MGM is stable in the total flux estimation of molecular clumps.

\begin{figure}[H]
	\centering
	\includegraphics[angle=0,width=86mm]{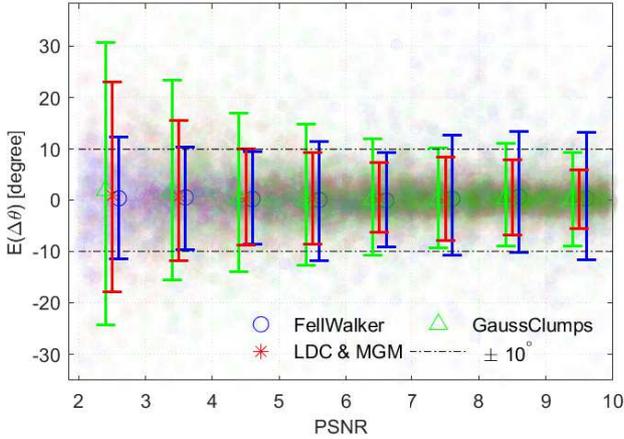}
	\caption{
		The statistics of absolute deviation in the tilt angle as a function of the PSNR. The tilt angle on the $ x-y $ plane of the molecular clump vary from $0 ^{\circ} $ to $180 ^{\circ} $. The minimum ratio of the major axis to the minor axis in these clumps is 1.4. The blue, green and red dots, special symbols and error bars have the same meaning as Figure~\ref{peak}.} 	
	\label{angle}
\end{figure}

Figure~$ \ref{angle} $ shows the deviation of tilt angle, the symbols are the same as Figure~\ref{peak}. We can see that the dispersions of the measured deviations are decreased gradually with increasing of the PSNR for the three algorithms, while the deviation is less than $10^{\circ}$ when the PSNR greater than 4, indicating that the estimation of molecular clump angle by this algorithm is stable.
\begin{figure*}[ht]
	\centering
	\includegraphics[angle=0,width=172mm]{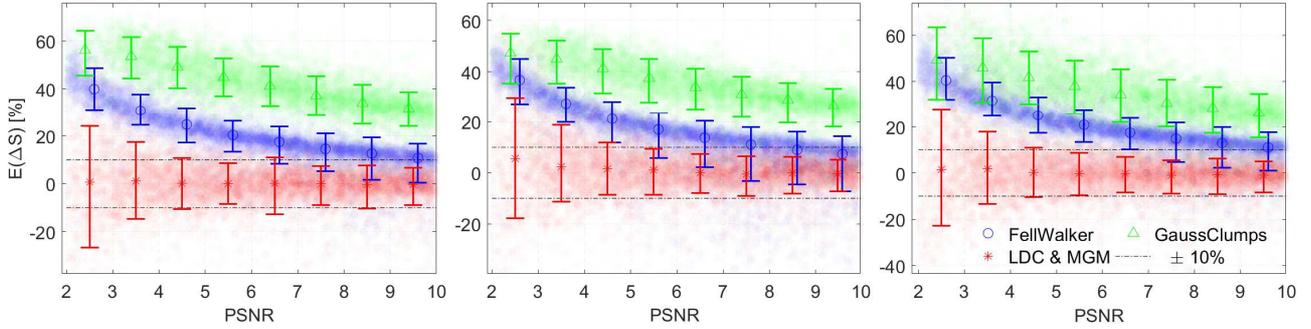}
	\caption{The statistics of relative deviation in size as a function of the PSNR. From left to right are the deviation of $ Size_1 $, $ Size_2 $, and $ Size_3 $, respectively ($ Size_1 $, $ Size_2 $ represent  major and minor size of detected clump in the spatial, $ Size_3 $ represents the size of detected clump in velocity axis). Two dashed horizontal lines represent the relative deviation of $ \pm10\% $. The blue, green and red dots, special symbols and error bars have the same meaning as Figure~\ref{peak}.} 
	\label{size}	
\end{figure*}
\begin{figure*}[ht]
	\centering
	\includegraphics[angle=0,width=172mm]{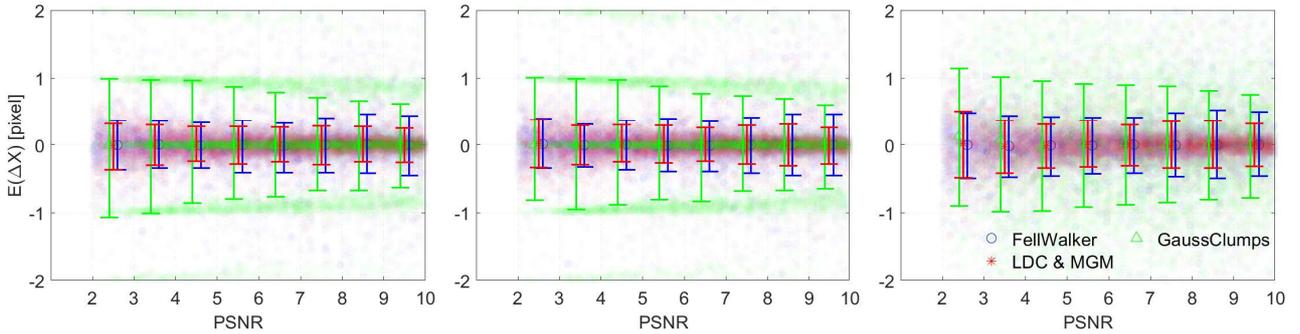}
	\caption{The statistics of absolute deviation of the position as a function of the PSNR. From left to right are the deviation of galactic latitude, galactic longititude, and velocity, respectively. The blue, green and red dots, special symbols and error bars have the same meaning as Figure~\ref{peak}.} 	
	\label{cen}	
\end{figure*}

The size of the molecular clump can be used to describe the different shapes of them, which is a very important parameter for the classification of the molecular clump. From left to right, the panels of Figure~$ \ref{size} $ show the statistics relative deviation in $ Size_1 $, $ Size_2 $, and $ Size_3 $, respectively. In Figure~$ \ref{size} $, we can see that the size obtained by GaussClumps exhibit a large deviation. The measured size of GaussClumps and FellWalker algorithms are lower than the simualted size. With increase of the PSNR, the deviations from the GaussClumps and FellWalker algorithms gradually decrease, while the deviations from the LDC \& MGM algorithm are closed to zeros. The deviation of LDC \& MGM is less than 10\% with the PSNR above 4, indicating that the size of clump obtained from the algorithm is reliable.

Figure~$ \ref{cen} $ shows the absolute deviation of position as a function of the PSNR, from left to right are the deviation on galactic latitude, galactic longititude, and velocity, respectively. The position deviations measured by FellWalker and LDC $ \& $ MGM are almost within 1 pixel and the deviation is no more than 0.5 pixel at the PSNR greater than 4, while the deviation from GaussClumps is greater than the two algorithms. We can see that some horizontal bars appear in the distribution of position measured by GaussClumps in galactic latitude and longititude direction. The reason is that the low spatial resolution of the simulated clumps leading to the centers fitted by Gaussclumps are mainly located on the grid.

Overall, detecting clumps by LDC \& MGM at high number density has robust parametration accuracy in term of position, peak, total flux, size, and tilt angle. The molecular clumps parametrization of the proposed algorithm show less deviation and less dispersion than FellWalker and GaussClumps algorithms with the PSNR above 5.

\section{Experiment in real data}
\label{sect:Experiment in real data}
\subsection{M16 Data}
\label{m16 introduction}
\begin{figure}[H]
	\centering
	\includegraphics[angle=0,width=86mm]{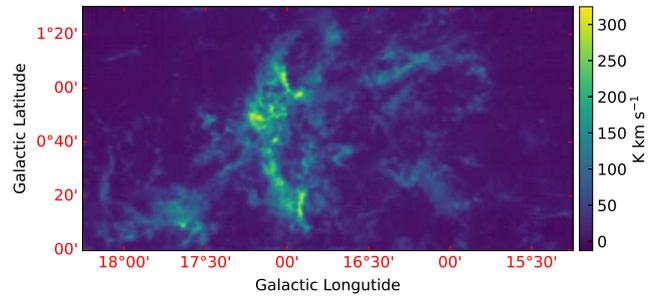}
	\caption{The intergrated intensity maps of M16 in $^{13}\rm CO$ (J=1-0) line with a volecity range of $ 15.93 - 27.06 $ km s$ ^{-1} $.} 	
	\label{M16_12_13_18_map}
\end{figure}

The $^{13}\rm CO$ (J=1-0) line of M16, including the region within 15$ ^{\circ} $15$^{\rm{'}}$ $ \textless$ $ l $ $ \textless $ 18$ ^{\circ} $15$^{\rm{'}}$ and 0$ ^{\circ} $ $ \textless $ $ |b| $ $ \textless $ 1$ ^{\circ} $30$^{\rm{'}}$ from the Milky Way Imaging Scroll Painting (MWISP) survey~\citep{2018AcASn..59....3S}, is employed in the molecular clump detection and parametrization. The typical noise level at $^{13}\rm CO$ (J=1-0) line is about 0.23 K with the channel width of 0.167 km s$ ^{-1} $. Figure~\ref{M16_12_13_18_map} shows the intergrated intensity maps of M16 in $^{13}\rm CO$ (J=1-0) line.

Using the M16 data, \cite{2016RAA....16...56Z} has confirmed the identification of the giant molecular filament
(GMF) G18.0-16.8 by \cite{2014A&A...568A..73R} and find a new giant filament, G16.5-15.8, located in the west 0.8$ {} ^{\circ}$ of G18.0-16.8. \cite{songchao} has calculated the properties of the clump samples under local thermodynamic equilibrium (LTE) assumption. The virial mass and virial parameter are calculated to evaluate whether clumps are bound or unbound. They found the majority of $^{13}\rm CO$ clumps are bound, which suggest that those clumps may form stars in the future. Based on their research in detection clump on M16, the $^{13}\rm CO$ (J=1-0) line of M16 is used to investigate the performance of our algorithm.
\subsection{Clump extraction experiment}
\label{Extracting parameters in the real data}
\begin{figure}[H]
	\centering
	\includegraphics[angle=0,width=86mm]{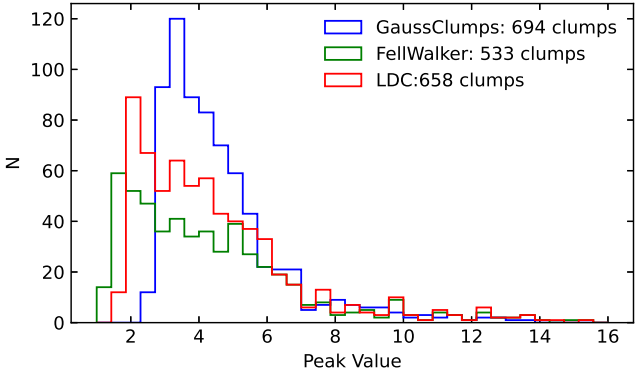}%"scale"后的数字为图形的宽度，也可用"width=1.0\columnwidth"定义

	\caption{
		The distribution of the detected peak intensity values of clumps by GaussClumps, FellWalker, and LDC.
	} 	
	\label{peak_distribution}
\end{figure}

After tuning the algorithm parameters, the GaussClumps, FellWalker and LDC algorithms are applied to detect the $^{13}\rm CO$ (J=1-0) line of M16. Figure~\ref{peak_distribution} shows the distribution of peak intensity value of clumps detected by the three algorithms. The observed total flux is defined as the summed flux of those observations above 2$ \times rms $ of the background. And the recovery rate is defined as the ratio the sum of clumps flux to the observed total flux. The recovery rate of total flux obtained by GaussClumps, FellWalker and LDC are 51.6\%, 90.4\% and 90.2\% in $^{13}\rm CO$ emission of M16, respectively.
It is can be infered from Figure~\ref{peak_distribution} that the peak intensity values of clumps detected by LDC and FellWalker have a similar distribution with a more flatted peak, while the distribution of the peak intensity values detected by GaussClumps deviates greatly from the other two algorithms. The peak of distributon is about 2 in the FellWalker and LDC algorithm, while the GaussClumps is for 3.4. Combined with the minimum peak intensity value (about 2.1 K) of clumps detected by GaussClumps and the noise level (0.23 K) at $^{13}\rm CO$ (J=1-0) line, it shows that the PSNR of clumps detected by GaussClumps are greater than 9, while the recall rate of the algorithm in the Section~\ref{Testing performance evaluation} can be maintained a certain level with the PSNR above 5. It may be the Gaussclumps algorithm tends to fit a clump with a strict elliptic shape, and it fails to fit a clump with weaker peak intensity value in the real data.

\subsection{Completeness}
\label{subsec completeness}
The limitation of the telescope sensitivity causes low quality clump being ignored. Other indicators of the algorithm are the completeness and the detection rate above the limition. The ``completeness limit" here refers to the total flux or mass above which a clump can be detected at certain level with an algorithm. The smaller and weaker molecular clumps, the less likely they are to be detected.

We designed the dataset by randomly inserting simulated clumps into the $ {}^{13}\rm{CO} $ (J=1-0) line  of M16. The peak intensity value of those simulated clump take values from 2 to 5, while the size of the clump in the velocity axis take values from 2 to 4 and the size in the galactic longtitude and latitude axes take values from 0.5 to 2. The clumps detected by GaussClumps, FellWalker and LDC algorithm are matched with the simulated clumps. The number of clumps within each total flux interval are counted, the completeness and the average detection rate above the limitation is obtained.

\begin{figure}[H]
	\centering
	\includegraphics[angle=0,width=86mm]{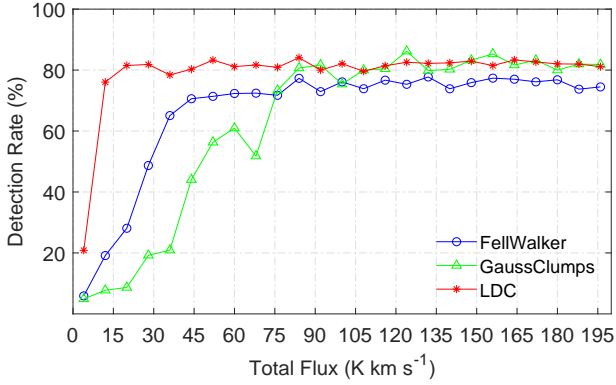}
	\caption{The detection rate of the three algorithms in $^{13}\rm CO$ (J=1-0) line of M16 as a function of total flux. The detection rate of GaussClumps, FellWalker and LDC are 80.9\%, 74.7\% and 81.7\%, respectively. The completeness limitation of LDC and FellWalker are 20 K km s$ ^{-1}$ and 45 K km s$ ^{-1}$, respectively. While the GaussClumps is 75 K km s$ ^{-1}$.} 	
	\label{completeness}
\end{figure}

\begin{figure}[H]
	\centering
	\includegraphics[angle=0,width=86mm]{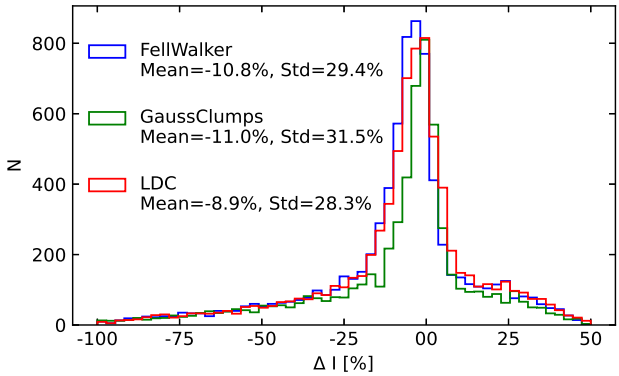}
	\caption{
		The histogram of the peak deviation ($ \Delta \rm{I} $) for the GaussClumps, FellWalker, and LDC. The $ \Delta \rm{I} $ is described in Formula~(\ref{d_peak}). The mean deviation of FellWalker, GaussClumps and LDC are -10.8\%, -11.0\% and -8.9\%, respectively. The standard deviation of FellWalker, GaussClumps and LDC are 29.4\%, 31.5\% and 28.3\%, respectively.
	} 	
	\label{completeness_peak}
\end{figure}

Figure~\ref{completeness} shows the detection rate of the GaussClumps, FellWlaker, and LDC algorithm in $^{13}\rm CO$ (J=1-0) line of M16, respectively. As the total flux increases, the detection rate of the GaussClumps grows slowly, while the Fellwalker and LDC are able to maintain a relatively high detection rate all the way from the completeness limitation. The detection rate of GaussClumps, FellWalker and LDC are 80.9\%, 74.7\% and 81.7\% above the completeness limitation, respectively. From the detection rate of each algorithm, we can roughly estimate that the number density in $ {}^{13}\rm{CO} $ (J=1-0) line of M16 is between the high density and medium density in the simulation datasets described in Section~\ref{3d simulation data}.
Figure~\ref{completeness_peak} shows the statistical histogram of $ \Delta \rm{I} $ for the three algorithms. The simulated clumps could overlap in real observations, leading to the detected peak intensity values of clumps by the three algorithms are systematically larger than those of simulated clumps. While the LDC has the least dispersion of deviations. The long tail at the left side of the peak deviation suggests a relatively high intensity of molecular clumps in M16.

\section{Conclusion}
\label{conclusion}
We present a molecular clump detection and parametrization algorithm based on the Local Density Clustering and Multiple Gaussian Model (LDC \& MGM). The proposed algorithm is robust and universal in the clump detection. The employed algorithm of LDC in the clump detection and segmentation could achieve high accuracy with different signal-to-noise levels, while the MGM could obtain reliable physical parameters of overlapping clumps.

We applied our method to a simulated data set, and find, (1) detection rate: the recall rate of the algorithm at high, medium and low number density simulated data is greater than 80\%, 90\%, and 97\% with the PSNR above 6, respectively. The algorithm retains a high level of detection accurary when the PSNR is greater than 3. (2) Accuracy of parameters: the parametrization of the algorithm in simulated data show less deviation and less dispersion with the PSNR above 5. The deviations of peak value and size are almost within 10\% with the PSNR above 5, while the deviations of total flux hardly exceed 30\% when the PSNR is greater than 4 at the high number density. The deviations of titl angle on the $ x-y $ plane are less than 10$ {}^{\circ} $ with the PSNR above 4.

We apply our algorithm to the $^{13}\rm CO$ (J=1-0) map of the M16 nebula taken by PMO-13.7m telescope. The detection rate of clumps is up to 81.7\% with a completeness limitation of 20 K km s$ ^{-1} $ in $^{13}\rm CO$ (J=1-0) line of M16. A total of 658 molecular clumps have been detected by our algorithm and the total flux recovery rate in  $^{13}\rm CO$ (J=1-0) line of M16 is estimated as 90.2\%. The number density in  $^{13}\rm CO$ (J=1-0) line of M16 may be between the high and medium density in the simulation datasets described in Section~\ref{3d simulation data}.

\normalem
\begin{acknowledgements}
We thank the anonymous referee for his/her suggestive comments that help improve the manuscript a lot. This work was supported by the National Natural Science Foundation of China (U2031202, 11903083, 11873093). This research made use of the data from the MWISP project, which is a multi-line survey in $^{12}\rm CO$/$^{13}\rm CO$/$\rm C^{18}O$ along the northern galactic plane with PMO-13.7m telescope. We are grateful to all the members of the MWISP working group, particulally the staff mambers at PMO-13.7m telescope, for their long-term support. MWISP was sponsored by National Key R\&D Program of China with grant 2017YFA0402701 and  CAS Key Research Program of Frontier Sciences with grant QYZDJ-SSW-SLH047.

\end{acknowledgements}
  
\bibliographystyle{raa}
\bibliography{bibtex}{}

\end{multicols}{}
\end{document}